\documentclass{article}
\usepackage[T1]{fontenc} 
\usepackage[utf8]{inputenc} 
\usepackage{ismir,amsmath,cite,url}
\usepackage{graphicx}
\usepackage{color}

\usepackage{amssymb}
\usepackage{bbm}

\usepackage{lineno}

\title{Improved Handling of Repeats and Jumps in Audio--Sheet Image Synchronization}

\twoauthors
  {Mengyi Shan} {Harvey Mudd College \\ {\tt mshan@g.hmc.edu}}
  {TJ Tsai} {Harvey Mudd College \\ {\tt ttsai@g.hmc.eud}}

\begin{document}

\maketitle
\begin{abstract}
This paper studies the problem of automatically generating piano score following videos given an audio recording and raw sheet music images.  Whereas previous works focus on synthetic sheet music where the data has been cleaned and preprocessed, we instead focus on developing a system that can cope with the messiness of raw, unprocessed sheet music PDFs from IMSLP.  We investigate how well existing systems cope with real scanned sheet music, filler pages and unrelated pieces or movements, and discontinuities due to jumps and repeats.  We find that a significant bottleneck in system performance is handling jumps and repeats correctly.  In particular, we find that a previously proposed Jump DTW algorithm does not perform robustly when jump locations are unknown a priori.  We propose a novel alignment algorithm called Hierarchical DTW that can handle jumps and repeats even when jump locations are not known.  It first performs alignment at the feature level on each sheet music line, and then performs a second alignment at the segment level.  By operating at the segment level, it is able to encode domain knowledge about how likely a particular jump is.  Through carefully controlled experiments on unprocessed sheet music PDFs from IMSLP, we show that Hierarachical DTW significantly outperforms Jump DTW in handling various types of jumps.
\end{abstract}
\section{Introduction}
\label{sec:introduction}

This paper tackles the problem of generating piano score following videos in a fully automated manner.  Given an audio recording of a piano performance, our long-term goal is to build a system that can (a) identify the piece and automatically download the corresponding sheet music PDF from the International Music Score Library Project (IMSLP) website, and (b) generate a video showing the corresponding line of sheet music at each time instant in the audio recording.  In this work, we focus exclusively on task (b), assuming that the correct sheet music PDF has been identified.  This task requires us to perform audio--sheet music alignment on completely unprocessed PDF files from IMSLP.  This paper describes the key insights we have gained in building such a system, along with a novel alignment algorithm developed in the process.

Many previous works have studied cross-modal alignment between sheet music images and audio.  Two general categories of approaches have been proposed.  The first approach is to convert the sheet music images to a symbolic representation using optical music recognition (OMR), to collapse the pitch information across octaves to get a chroma representation, and then to compare this representation to chroma features extracted from the audio.  This approach has been applied to synchronizing audio and sheet music \cite{DammFKMC08_MultimodalPresentationofMusic_ICMI}\cite{KurthMFCC07_AutomatedSynchronization_ISMIR}\cite{ThomasFMC12_LinkingSheetMusicAudio_DagstuhlFU}, identifying audio recordings that correspond to a given sheet music representation \cite{FremereyMKC08_AutomaticMapping_ISMIR}, and finding the corresponding audio segment given a short segment of sheet music \cite{FremereyCME09_SheetMusicID_ISMIR}.  The second approach is to convert both sheet music and audio into a learned feature space that directly encodes semantic similarity.  This has been done using convolutional neural networks combined with canonical correlation analysis \cite{dorfer2016towardsEnd}\cite{dorfer2018end}, pairwise ranking loss \cite{dorfer2017learning}\cite{dorfer2018tismir}, or some other suitable loss metric.  This approach has been explored in the context of online sheet music score following \cite{dorfer2016live}, sheet music retrieval given an audio query \cite{dorfer2016towards}\cite{dorfer2017learning}\cite{dorfer2018tismir}, and offline alignment of sheet music and audio \cite{dorfer2017learning}.  Recent works \cite{dorfer2018learningToListen}\cite{henkel2019score} have also shown promising results formulating the score following problem as a reinforcement learning game.  See \cite{mueller2019cross} for an overview of work in this area.

The main difference between our current task and previous work is that we are working with totally unprocessed data ``in the wild."  All of the above works make one or more of the following assumptions, which are untrue in our task.  First, many works focus primarily on training and testing with synthetic sheet music.  In our case, we are primarily working with digital scans of physical sheet music.  Second, most works assume that the data has been cleaned and preprocessed in various ways.  For example, it is commonly assumed that unrelated pages of sheet music have been removed.  Many works further assume that each page has been segmented into lines, so that the data is presented as a sequence of image strips each containing a single line of sheet music.  In our task, the raw PDF from IMSLP may contain unrelated movements, pieces, or filler pages like the title page or table of contents.  We also obviously cannot assume that each page has already been segmented perfectly.  Third, all of the above works assume that the music does not have any jumps or repeats.  In our task, we have to be able to handle common discontinuities like repeats, D.C. al coda, D.S. al fine, etc.

In attempting to build a system that can handle messy, real-world data, we discovered two things.  First, we found that most of the above issues can be resolved to a reasonable degree by suitably combining existing tools in the MIR literature.  However, we also discovered that a significant bottleneck in system performance was handling jumps and repeats.  In particular, we found that a previously proposed Jump DTW alignment algorithm \cite{FremereyMC10_RepeatsJumps_ISMIR} does not yield satisfactory performance when jump locations are unknown a priori.

\begin{figure}
	\centerline{
		\includegraphics[width=\columnwidth]{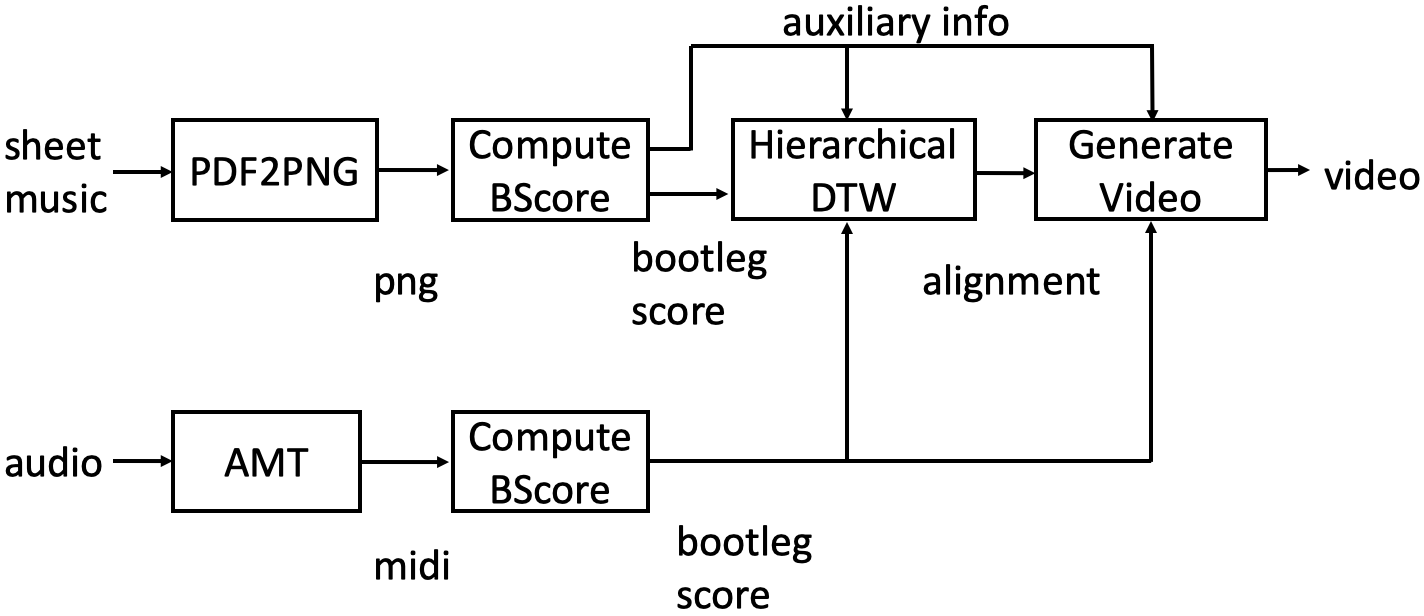}}
	\caption{Architecture of proposed system.  The sheet music and audio are both converted into bootleg scores, and then aligned with the Hierarchical DTW algorithm.}
	\label{fig:systemBlockDiagram}
\end{figure}

There are several existing offline algorithms for aligning two feature sequences in the presence of jumps or repeats.  Jump DTW \cite{FremereyMC10_RepeatsJumps_ISMIR} is a variant of dynamic time warping (DTW) where additional long-range transitions are allowed in the cost matrix at potential jump locations.  Mueller and Appelt \cite{MuellerA08_PathConstrained_ICASSP} and Grachten et al. \cite{grachten2013automatic} also propose variants of DTW for partial alignment in the presence of structural differences.  One limitation of these latter two works is that repeated sections are handled by simply skipping or deleting sections of features, so that the actual alignment of the repeated section is not known.  Joder et al. \cite{joder2011conditional} frame the alignment problem as a conditional random field with additional transitions inserted at known jump locations.  Jiang et al. \cite{jiang2019offline} use a modified Markov model that allows arbitrary jumps to follow a musician during a practice session with lots of do-overs and jumps.  There are also several real-time score following algorithms that handle various types of jumps \cite{nakamura2015real}\cite{arzt2010towards}\cite{arzt2008automatic}\cite{pardo2005modeling}, though our focus in this work is on the offline context.  In this study, we primarily focus on Jump DTW as the closest match to our target scenario: it is an offline algorithm, targeted at performances rather than practice sessions, and it provides a complete estimated alignment in the presence of jumps.

The main conceptual contribution of this paper is a novel alignment algorithm called Hierarchical DTW.  Unlike Jump DTW, it does not require knowledge of jump locations a priori, but instead considers every line transition as a potential jump location.  The algorithm is called Hierarchical DTW because it first performs an alignment at the feature level with each sheet music line, and then uses the results to perform a second alignment at the segment level.  By performing an alignment at the segment level, we can encode domain knowledge about which types of jumps are likely.  The algorithm is very simple and only has two hyperparameters, which both have very clear and intuitive interpretations.  Through carefully controlled experiments on unprocessed PDFs from IMSLP, we show that Hierarchical DTW significantly outperforms Jump DTW on the piano score following video generation task.\footnote{Code, data, and example score following videos can be found at \url{https://github.com/HMC-MIR/YoutubeScoreFollowing}.}


\section{System Description}
\label{sec:systemDescr}

Figure \ref{fig:systemBlockDiagram} shows a high-level overview of our proposed system.  We will explain its design in three parts: feature extraction, alignment, and video generation.

\subsection{Feature Extraction}

The first step is to convert both the sheet music and audio into bootleg score representations.  The bootleg score \cite{yang2019midipassage} is a recently proposed feature representation for aligning piano sheet music images and MIDI.  For sheet music, it encodes the position of filled noteheads relative to the staff lines.  The bootleg score itself is a $62 \times N$ binary matrix, where 62 indicates the total number of possible staff line positions in both the left and right hands, and where $N$ indicates the total estimated number of simultaneous note events.  For MIDI files, each note onset can be projected onto the bootleg score using the rules of Western musical notation.  Ambiguities due to enharmonic representations or left-right hand attribution are handled by simply setting all possible positions to 1.

We computed the bootleg score representations in the following manner.  We convert each PDF into a sequence of PNG images at 300 dpi, compute a bootleg score for each page, and then represent the entire PDF as a sequence of bootleg score fragments, where each fragment corresponds to a single line of music.  Note that these fragments may include lines of music from other unrelated movements or pieces in the same PDF, or may even represent nonsense features coming from filler pages.  Next, we transcribe the audio recording using the Onsets and Frames \cite{hawthorne2017onsets} automatic music transcription system, and then convert the estimated MIDI into its corresponding bootleg score.  In this work, we treat the bootleg score computation and music transcription as fixed feature extractors.

\subsection{Alignment}

\begin{figure}
	\centerline{
		\includegraphics[width=\columnwidth]{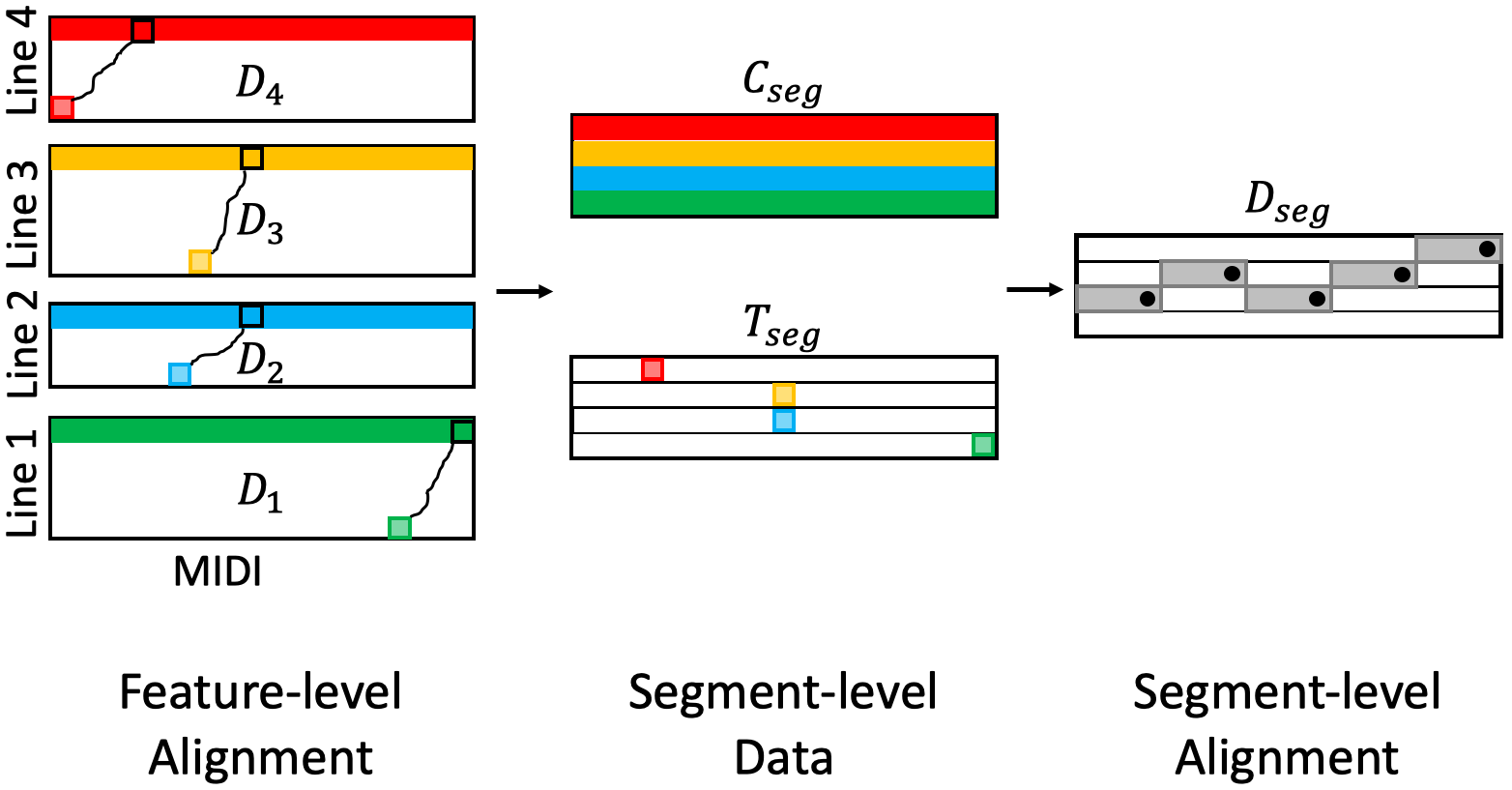}}
	\caption{Overview of Hierarchical DTW.  Subsequence DTW is performed at the feature level on each sheet music line.  The results are used to generate the segment-level data matrices, and then a second alignment is performed at the segment level.  Only a few selected elements of $T_{seg}$ are shown for illustration.}
	\label{fig:hierarchicalDTW}
\end{figure}

The second main step is to align the bootleg score representations.  We propose a novel alignment algorithm called Hierarchical DTW to accomplish this task.  Figure \ref{fig:hierarchicalDTW} shows an overview of the algorithm, which consists of three stages.

The first stage is to perform feature-level alignment.  We do this using a variant of DTW called subsequence DTW, which finds the optimal alignment between a short query sequence and any subsequence within a reference sequence.  We perform subsequence DTW between each sheet music bootleg score fragment (each corresponding to one line of music) and the entire MIDI bootleg score, as shown on the left side of Figure \ref{fig:hierarchicalDTW}.\footnote{In Figure \ref{fig:hierarchicalDTW}, the horizontal axis corresponds to the reference (left to right) and the vertical axis corresponds to the query (bottom to top).}  We use the normalized negative inner product distance metric proposed in \cite{yang2019midipassage} along with allowable transitions $\{(1,1), (1,2), (2,1)\}$ with weights $\{1, 1, 2\}$.  For a more detailed explanation of subsequence DTW, we refer the reader to \cite{muller2015fundamentals}.

The second stage is to construct the segment-level data matrices.  There are two matrices that need to be constructed.  The first matrix is formed by taking the last row in every cumulative cost matrix $D_i$ from stage 1 and stacking them into a matrix of size $L \times M$, where $L$ indicates the total number of lines of music in the PDF and $M$ indicates the total number of features in the MIDI bootleg score.  This matrix contains subsequence path scores and is denoted as $C_{seg}$ in Figure \ref{fig:hierarchicalDTW}.  It will play a role analogous to the pairwise cost matrix when we do dynamic programming at the segment level.  The second matrix $T_{seg}$ is the same size as $C_{seg}$ and indicates allowable transitions at the segment level.  Each element $T_{seg}[i,j]$ is computed by identifying the $j^{th}$ element in the last row of $D_i$, and then backtracking from this element to determine the beginning location of the matching path.  $T_{seg}[i,j]$ thus indicates the starting location of the best matching path in the $i^{th}$ line of sheet music ending at position $j$ in the MIDI bootleg score.  In Figure \ref{fig:hierarchicalDTW}, a few selected elements in $T_{seg}$ are shown as colored boxes to illustrate this process.  Note that, in order to construct $T_{seg}$, we need to backtrace from every possible location for every line of sheet music.

The third stage is to perform segment-level alignment.  Here, we use dynamic programming to find the optimal path through $C_{seg}$ using transitions in $T_{seg}$.  We construct a segment-level cumulative cost matrix $D_{seg}$ by filling out its entries column-by-column using dynamic programming.  The first column of $D_{seg}$ is initialized to all zeros, which ensures that the matching path can start on any line of music without penalty.  Note that, unlike regular DTW where the set of allowable transitions and weights is the same at every location, here the set of allowable transitions and weights is different for each element of $D_{seg}$.  Since the transitions are all unique, we simply encode the previous location rather than the transition type (e.g. the previous location $(i-1, j-1)$ instead of the transition $(1,1)$).  When computing $D_{seg}[i,j]$, there are two types of allowable transitions.  The first type of transition is skipping elements.  This means transitioning from $(i, j-1)$ and moving directly to the right by one position without accumulating any score.  Here, the candidate path score is $D_{seg}[i,j] = D_{seg}[i, j-1]$.  The second type of transition is matching the $i^{th}$ line of music (ending) at this position.  In this case, we can transition from the end of any line of music immediately before the matching segment begins.  If we let $k \triangleq T_{seg}[i,j]$ be the beginning of the matching subsequence path, then there are $L$ different possible transitions from $(n, k-1), n=0,\dots, L-1$ where $n$ indicates the line of music.  Here, the candidate path scores are $D_{seg}[i,j] = D_{seg}[n, k-1] + w_{n, i} \cdot C_{seg}[i, j] + p_{n,i}$, where $w_{n,i}$ is a multiplicative weight and $p_{n,i}$ is an additive penalty for jumps.  We can summarize the dynamic programming rules for the segment-level alignment as
\begin{align*}
k &\triangleq T_{seg}[i,j] \\
D_{seg}[i,j] &= min
\begin{cases}
D_{seg}[i, j-1] \\
D_{seg}[0, k-1] + w_{0,i} \cdot C_{seg}[i,j] + p_{0,i} \\
D_{seg}[1, k-1] + w_{1,i} \cdot C_{seg}[i,j] + p_{1,i} \\
\cdots \\
\end{cases}
\end{align*}
where the minimum is calculated over all sheet music lines $n=0,\dots,L-1$.  When filling out the entries of $D_{seg}$ using dynamic programming, we also keep track of backtrace information in a separate matrix.  Once $D_{seg}$ has been constructed, we identify the element in the last column of $D_{seg}$ with the lowest path score, and then backtrace from that position to determine the optimal alignment path.  Figure \ref{fig:hierarchicalDTW} shows the optimal alignment path as a series of black dots and the induced segmentation of the MIDI bootleg score as gray rectangles.

The real power of Hierarchical DTW comes from setting $w_{n,i}$ and $p_{n,i}$ in an intelligent way that encodes musical domain knowledge.  These values can be adapted to allow no jumps, allow arbitrary jumps, or anything in between.  For example, disallowing jumps means setting $p_{n,i} = \infty \cdot \mathbbm{1}(i \ne n+1)$.  The system described below is one possible instantiation based on three assumptions: (a) the performed lines of music will form a contiguous block (e.g. we will not go from page 13 to 34 to 19), (b) backwards jumps (from repeats) are to lines of music we have seen before, and (c) forward jumps (from D.S. al fine) are to one line past the furthest line of music that has been seen before (which we refer to as the ``leading edge").  For the allowed jump transitions, multiplicative weights are set to $1$ and additive penalties are set to $-\gamma \cdot p_{avg}$, where $\gamma$ is a hyperparameter and $p_{avg}$ is the result of calculating the best subsequence path score for each line of sheet music and averaging the scores across all lines.  So, if $\gamma = 1$, the jump penalty approximately offsets 1 line of matching music.  Note that we can keep track of which lines have been seen before by defining two matrices $R_{lower}$ and $R_{upper}$ which are the same size as $C_{seg}$ and keep track of the range of lines that have been seen in the optimal path ending at any position $(i,j)$.  $R_{lower}$ and $R_{upper}$ can be updated along with $D_{seg}$ and the backtrace matrix during the dynamic programming stage.  For regular forward transitions, we allow moving to the next line, staying on the current line (slowing down), or skipping one line (speeding up).  These three transitions have multiplicative weights $1$, $\alpha$, and $\alpha$ and additive penalties of $0$ (all), respectively.  We found that allowing additional time warping at the segment level with multiplicative penalty $\alpha = 0.5$ allows the algorithm to recover from large mistakes more easily.  

Hierarchical DTW is simple yet flexible.  The version described above only has two hyperparameters that correspond to a multiplicative penalty for speeding up/slowing down ($\alpha$) and an additive penalty for jumps ($\gamma$).  Yet, the framework of Hierarchical DTW makes it possible to selectively allow very specific types of jumps that obey common musical conventions.

\subsection{Video Generation}

The third main step is to generate the score following video.  In order to translate the predicted segment-level alignment into a score following video, we need additional auxiliary information from the bootleg score feature computation.  For the audio recording, we need to keep track of the correspondence between each MIDI bootleg score feature column and its corresponding time in the audio recording.  For the sheet music, we need to keep track of the correspondence between each sheet music bootleg score feature column and its corresponding page and pixel range in the sheet music images.  We modified the original code provided in \cite{yang2019midipassage} to return this information, in addition to the bootleg score features.  Given this auxiliary information and the predicted segment-level alignment, we can generate the score following video in a very straightforward manner: we simply show the predicted line of sheet music at every time instant in the audio recording.

\section{Experimental Setup}
\label{sec:setup}

In this section, we explain the datasets and metrics used to evaluate our proposed system.

\begin{figure}
	\centerline{
		\includegraphics[width=\columnwidth]{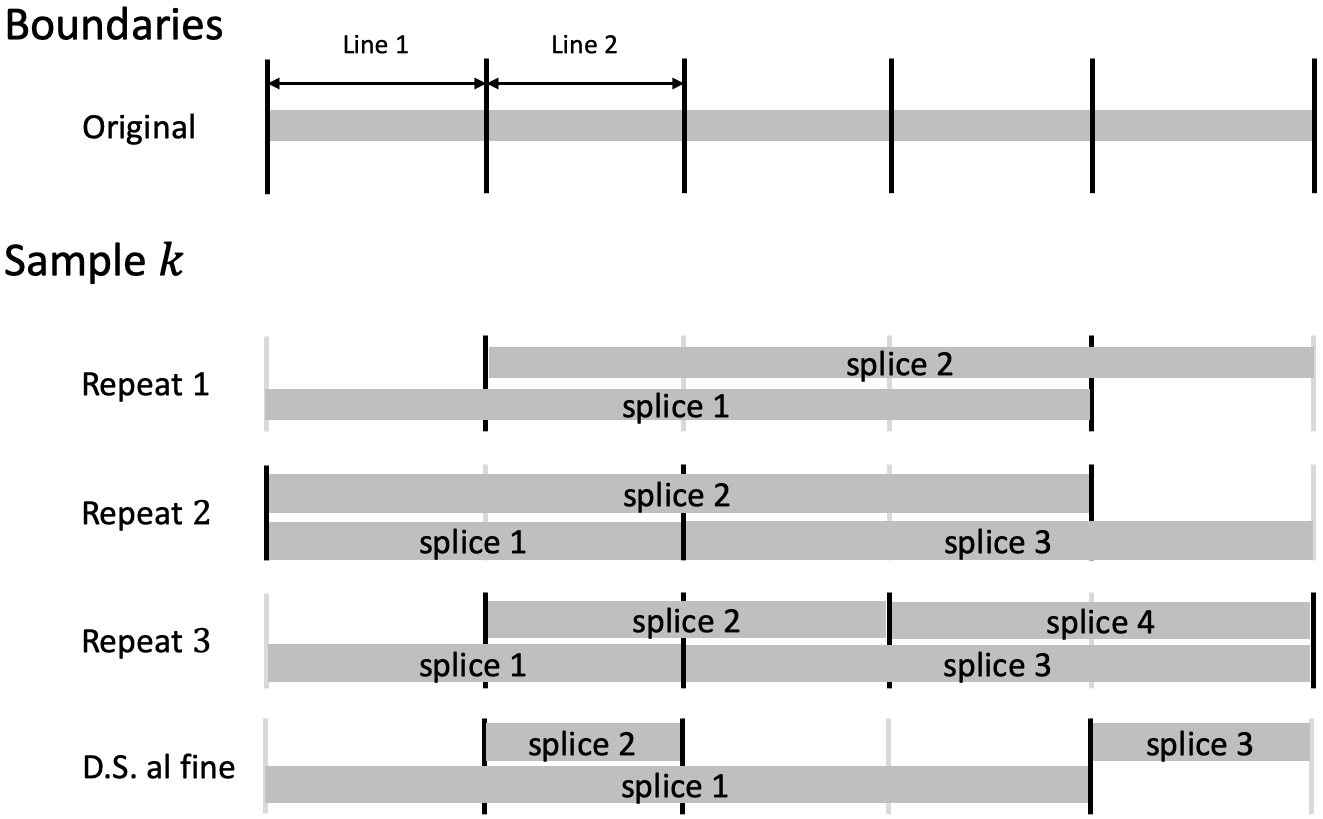}}
	\caption{Generating audio with repeats.  The original audio recording is segmented by lines of sheet music.  We sample $k$ boundary points without replacement, and then splice and concatenate audio segments to generate the data with repeats.}
	\label{fig:generatingData}
\end{figure}

Our data is a derivative of the Sheet MIDI Retrieval dataset \cite{yang2019midipassage}.  We will first describe the original dataset, and then explain how we used it to generate the data for this current work.  The original dataset contains scanned sheet music from IMSLP for 200 solo piano pieces across 25 composers.  The sheet music comes with manual annotations of how many lines of music are on each page, and how many measures are on each line.  For each of the 200 pieces, there is a corresponding MIDI file and ground truth annotations of measure-level timestamps.

We derived our dataset in the following manner.  We synthesize the MIDI files to audio using the FluidSynth library.  By combining the sheet music and MIDI annotations, we determine the time intervals in the audio recording that correspond to each line of sheet music.  For each sheet music PDF in the Sheet MIDI Retrieval dataset, we retrieved the original PDF from the IMSLP website.  The only difference between these two files is that the original IMSLP PDF contains other unrelated movements, pieces, and filler pages that were removed during the preparation of the Sheet MIDI Retrieval dataset.  For example, one PDF in the test set contains 127 pages, of which only 17 correspond to the piece of interest.  Because we want to test how well our system handles this type of noise, we use the original PDF with no preprocessing or data cleaning whatsoever.  We augmented the sheet music annotations by converting the original IMSLP PDFs into PNG files at 300 dpi and manually annotating the vertical pixel range for every line of sheet music played in the audio recording.  This required annotating a total of $1090$ pages with $11,556$ pixel positions.  By combining all of our annotations together, we can determine the page and pixel range of the line of sheet music that is currently being played at every point in the audio recording.  In total, there are $13.0$ hours of annotated audio.  Because there are no repeats or jumps in the sheet music, we call this data the ``No Repeat" dataset.

We also generate several synthetic datasets to test how well our system handles jumps and repeats.  The process of generating a synthetic dataset consists of three steps, as shown in Figure \ref{fig:generatingData}.  The first step is to identify the $L+1$ boundary positions of the $L$ lines of sheet music that are played in the audio recording.  The second step is to randomly sample $k$ boundary points without replacement.  The value of $k$ depends on the types of jumps we want to simulate.  In this work, we consider four schemas: 1 repeat ($k=2$), 2 repeats ($k=3$), 3 repeats ($k=4$), and D.S. al fine ($k=3$).  The third step is to splice and concatenate the audio to generate a modified audio recording as shown in Figure \ref{fig:generatingData}.  Note that all of the synthetic datasets have the exact same sheet music, but their audio recordings have been spliced to reflect the desired schema.  Since the process of sampling is random, we generate five different samples for every audio recording.  The four synthetic datasets described above have $84$, $94$, $100$, and $81$ hours of audio, respectively.  The ground truth annotations are modified accordingly.

We evaluate system performance using a simple accuracy metric.  Because our goal is to generate score following videos, we want to use an evaluation metric that correlates with user experience.  The accuracy simply indicates the percentage of time that the correct line of music is being shown to the user.  When calculating accuracy, we use a scoring collar, in which small intervals $(t_i - \Delta t, t_i + \Delta t)$ around the ground truth transition timestamps $t_i$ are ignored during scoring.  This is a standard practice in evaluating time-based segmentation tasks like speech activity detection \cite{nist2016sadeval}.  By using a range of scoring collar values, we can also gain insight into what fraction of our errors occur very close to the transition boundaries.

For all experiments, we use (the same) 40 pieces for training and 160 pieces for testing.  This results in 160 test queries for the No Repeat benchmark ($10.6$ hours of audio) and $160 \times 5 = 800$ test queries for the benchmarks with jumps ($69.2$, $76.9$, $81.8$, and $66.1$ hours).  Since we treat the bootleg score computation and automatic music transcription as fixed feature extractors, our system has no trainable weights and only 2 hyperparameters ($\alpha$, $\gamma$).  So, we only use a small fraction of the data for developing the algorithm, and we reserve most of the data for testing.

\section{Results}
\label{sec:results}

In this section, we present our experimental results on the piano score following video generation task.

\begin{figure}
	\centerline{
		\includegraphics[width=\columnwidth]{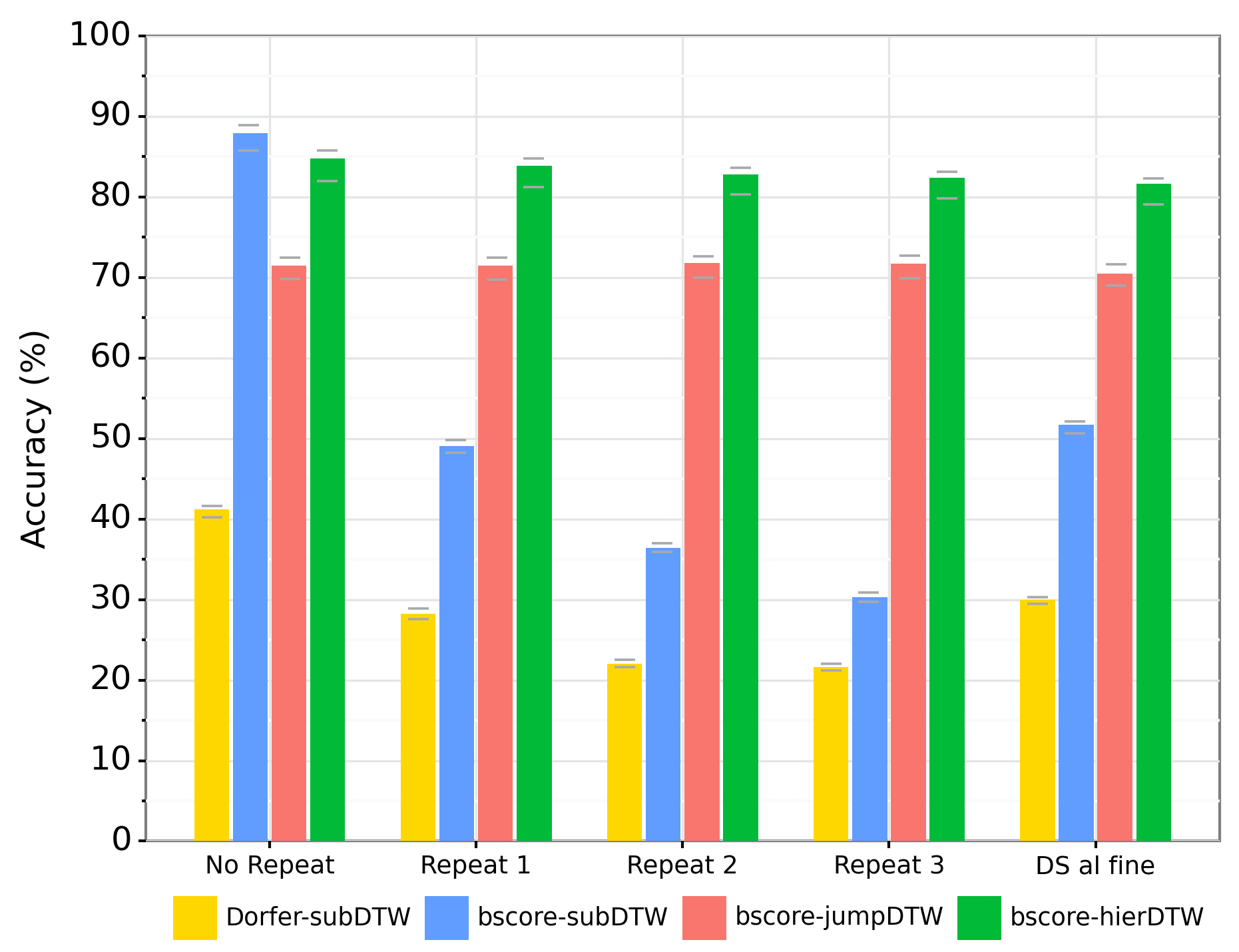}}
	\caption{Comparison of system performance on benchmarks with various types of jumps.  The bar levels indicate accuracy with a scoring collar of $0.5$ sec.  The short gray lines indicate accuracy with scoring collars of $0$ and $1.0$ seconds.}
	\label{fig:results}
\end{figure}

We compare our proposed system to three other baseline systems.  The first baseline system (`bscore-subDTW') is identical to our proposed system in Figure \ref{fig:systemBlockDiagram} except that it replaces the Hierarchical DTW with a simple subsequence DTW.  The second baseline system (`bscore-jumpDTW') is also identical to our proposed system except that it replaces the Hierarchical DTW with Jump DTW \cite{FremereyMC10_RepeatsJumps_ISMIR}.  Because Jump DTW was designed to handle jumps and repeats, we expect this system to provide the most competitive baseline results.  The third baseline system (`Dorfer-subDTW') is based on Dorfer et. al \cite{dorfer2018tismir}.  This system approaches the audio--sheet music alignment task by training a multimodal CNN to project chunks of sheet music and chunks of audio spectrogram into the same feature space where similarity can be computed directly.  We used the pretrained CNN provided in \cite{dorfer2018tismir} as a feature extractor, and then apply subsequence DTW.  Finally, our proposed Hierarchical DTW system is indicated as `bscore-hierDTW.'

Figure \ref{fig:results} shows the results of these four systems.  The histogram bars indicate the accuracies with a scoring collar of $\Delta t = .5$ sec.  There are four things to notice about these results.  First, the Dorfer-subDTW system performs poorly on all benchmarks.  This indicates that this system does not generalize well to the scanned sheet music from IMSLP.  Second, the bscore-subDTW system performs well on the No Repeat benchmark ($87.9\%$ accuracy), but performs poorly on all other benchmarks (e.g. $30.3\%$ on the Repeat 3 benchmark).  This is to be expected, since subsequence DTW cannot handle jumps and repeats.  Third, Jump DTW is significantly worse than subsequence DTW on the No Repeat benchmark ($71.5\%$  vs. $87.9\%$), but it has consistent performance across benchmarks with repeats and jumps ($71.5\%$, $71.8\%$, $71.7\%$, and $70.5\%$).  This indicates that Jump DTW is able to cope with discontinuities, but with a significant cost in performance.  Fourth, the Hierarchical DTW system is only slightly worse than subsequence DTW on the No Repeat benchmark ($84.8\%$ vs. $87.9\%$), and its performance decreases only slightly on the other benchmarks ($83.9\%$, $82.8\%$, $82.4\%$, $81.6\%$).  We can see that the Hierarchical DTW system consistently outperforms Jump DTW by 10-13\% across all benchmarks.  These results indicate that Hierarchical DTW is able to handle repeats and jumps reasonably well, and with a much smaller performance cost than Jump DTW.

\section{Analysis}
\label{sec:analysis}

In this section, we conduct two different analyses to gain more insight into system behavior.

\subsection{Failure Modes}

The first analysis answers the question, ``What are the failure modes for each system?"  To answer this question, we identified the individual queries that had the poorest accuracy, and then investigated the reasons for the errors.

The Dorfer system has two primary failure modes.  The first failure mode is that the system is not designed to handle jumps, so it performs very poorly on any datasets with jumps or repeats.  Note, however, that this system also performs poorly on the No Repeat benchmark.  When we investigated the reasons for this, we discovered the second major failure mode: page segmentation.  The sub-system for segmenting each page into lines of music performed very poorly on many pages in the dataset.  This is perhaps not surprising, since the original system was developed and trained on synthetic sheet music, where staff lines are perfectly horizontal.  In this case, the assumptions in this work do not translate well to our task of working with IMSLP scanned sheet music.

The subsequence DTW system also has two primary failure modes.  The first is (again) that the system cannot handle jumps or repeats.  When we investigated the reasons for major errors on the No Repeat benchmark, we find that the failures primarily come from mistakes in the bootleg score representation.  The bootleg score does not account for octave markings or clef changes, and it does not detect non-filled noteheads (e.g. half or whole notes).  When there are long stretches of sheet music that contain several of these elements at the same time, the bootleg score is a poor representation of the sheet music.  For example, three of the pieces in the test set are Erik Satie's Gymnopedies, where the sheet music is almost entirely non-filled noteheads.  These pieces had close to 0\% accuracy and caused a decrease of several percentage points on the aggregate accuracy score.

The JumpDTW system has one primary failure mode: it often jumps to incorrect lines of music.  This occurs when either (a) there are similar lines of music in multiple places (e.g. the recapitulation of a theme), or (b) significant bootleg score errors cause the system to match random lines of music elsewhere in the piece.  This problem is most clearly seen in the No Repeat benchmark, where it often takes jumps when none are present.

The Hierarchical DTW system has two primary failure modes.  The first failure mode is prolonged bootleg score failures, which cause the algorithm to insert spurious small jumps.  Once the bootleg score becomes an accurate representation again, the system is usually able to recover.  The second failure mode is when the sheet music contains very repetitive measures and lines.  This problem is particularly bad when the sheet music is very short (e.g. 2-3 pages long) and has jumps or repeats.

\begin{figure}
	\centerline{
		\includegraphics[width=\columnwidth]{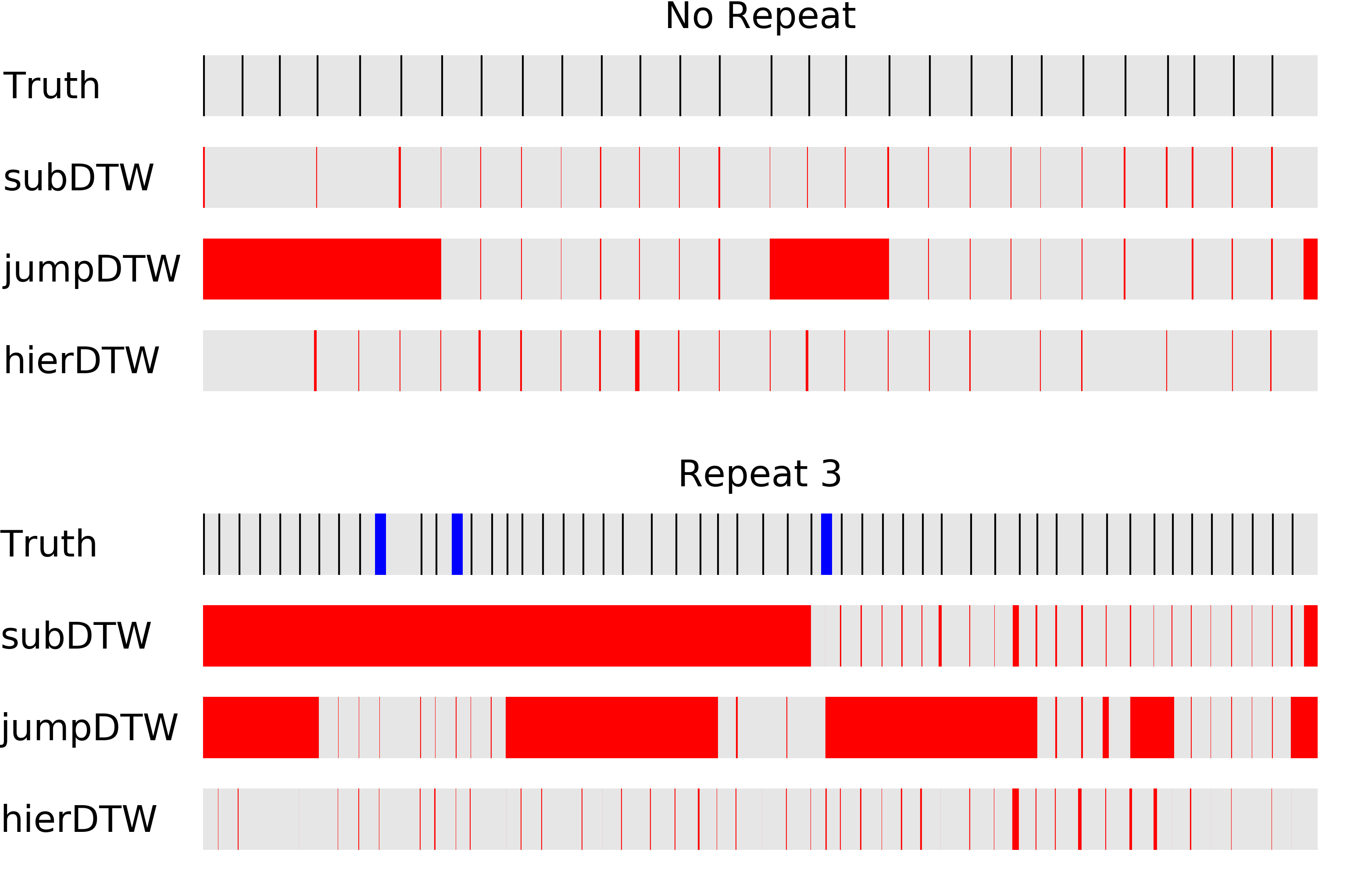}}
	\caption{Visualization of system predictions for a query with no repeats (top half) and a query with three repeats (bottom half).  Each gray strip indicates the duration of the audio recording.  The black vertical lines show the ground truth line transitions, and the red regions indicate times when an incorrect line of sheet music is being shown.}
	\label{fig:errorVisualization}
\end{figure}

Figure \ref{fig:errorVisualization} shows a visualization tool that was helpful in diagnosing failure modes.  The top half of Figure \ref{fig:errorVisualization} shows four gray strips, each representating the duration of a single audio recording in the No Repeat benchmark.  The topmost strip contains black vertical lines indicating the location of the ground truth sheet music line transitions.  The three strips below it show the predictions of the subsequence DTW, Jump DTW, and Hierarchical DTW systems, where errors are shown in red.  The bottom half of Figure \ref{fig:errorVisualization} shows the same information for a query in the Repeat 3 benchmark.  The location of the jumps are indicated with blue vertical lines.  We can see many of the failure modes described above.  For example, Jump DTW has spurious jumps in both queries but is able to follow two of the repeats in the bottom query.  Subsequence DTW is unable to handle the jumps in the bottom query, but matches well after the last jump occurs.  Finally, we can see that the Hierarchical DTW system is able to follow the correct sequence of sheet music lines, and its errors primarily occur close to line transitions.

\subsection{Error Locations}

The second analysis answers the question, ``Where are the errors located?"  One way we can answer this question is to calculate system performance across a range of values for the scoring collars.  This can tell us how close the errors are to line transition boundaries.

Figure \ref{fig:results} shows the results of each system with various scoring collar values.  The histogram bar level indicates the default scoring collar $\Delta t = .5$ sec, and the results with $\Delta t$ set to $0$ sec and $1.0$ sec are shown as short horizontal gray lines directly below and above the histogram bar level, respectively.  Note that as $\Delta t$ increases, the accuracy will increase monotonically.  

There are two things to notice about the results with various scoring collars.  First, we see that even with a generous scoring collar of $\Delta t = 1$ sec, the accuracies of all systems only increase about 1-2\%.  This indicates that most of the errors are not slight misalignments at the line transitions, but are instead large errors due to total alignment failures.  Second, we observe that the results with Hierarchical DTW on benchmarks with jumps is only marginally worse than the No Repeat benchmark.  This indicates that Hierarchical DTW is able to handle discontinuities reasonably well.  Combining these two observations, the failures in the bscore-hierDTW system seem to primarily come from large misalignments due to prolonged bootleg score failures.  This strongly suggests that the performance bottleneck is the bootleg score representation, not the Hierarchical DTW alignment.

\section{Conclusion}
\label{sec:conclusion}

We present a method for generating piano score following videos.  Our approach uses several recently proposed systems to convert both the sheet music and audio into bootleg score representations.  We then apply a novel alignment algorithm called Hierarchical DTW, which performs alignment at both the feature-level and the segment-level in order to handle repeats, jumps, and unknown offset in the sheet music.  We perform experiments with completely unprocessed sheet music from IMSLP, and we show that Hierarchical DTW significantly outperforms a previously proposed Jump DTW algorithm for handling jumps and repeats.  For future work, we would like to augment the system to automatically identify a piece and retrieve the corresponding sheet music from the IMSLP database in an automated fashion.

\bibliography{YoutubeScoreFollowing}

%
%
%
%

\end{document}